\documentclass[pdflatex,sn-mathphys-num]{sn-jnl}

\usepackage{graphicx}
\usepackage{multirow}
\usepackage{amsmath,amssymb,amsfonts}
\usepackage{amsthm}
\usepackage{mathrsfs}
\usepackage[title]{appendix}
\usepackage{xcolor}
\usepackage{textcomp}
\usepackage{manyfoot}
\usepackage{booktabs}
\usepackage{braket} 
\usepackage{algorithm}
\usepackage{algorithmicx}
\usepackage{algpseudocode}
\usepackage{listings}

\theoremstyle{thmstyleone}

\theoremstyle{thmstyletwo}

\theoremstyle{thmstylethree}

\begin{document}

\title[Architectural Impacts on Quantum Algorithms]{Investigation of Hardware Architecture Effects on Quantum Algorithm Performance: A Comparative Hardware Study}

\author*[1]{\fnm{Askar} \sur{Oralkhan}}\email{askaroralkhan@gmail.com}

\author[1]{\fnm{Temirlan} \sur{Zhaxalykov}}\email{t.zhaxalykov@kbtu.kz}

\affil*[1]{\orgdiv{School of Information Technology and Engineering}, \orgname{Kazakh British Technical University}, \orgaddress{\city{Almaty}, \country{Kazakhstan}}}

\abstract{
Cloud-accessible quantum processors enable direct execution of quantum algorithms on heterogeneous hardware platforms. Unlike classical systems, however, identical quantum circuits may exhibit substantially different behavior across devices due to architectural variations in qubit connectivity, gate fidelity, and coherence times.

In this work, we systematically benchmark five representative quantum algorithms—Bell state preparation, GHZ state generation, Quantum Fourier Transform (QFT), Grover’s Search, and the Quantum Approximate Optimization Algorithm (QAOA)—across trapped-ion, superconducting, and simulator backends using Amazon Braket. Performance metrics including fidelity, CHSH violation, success probability, circuit depth, and gate counts are evaluated.

Our results demonstrate a strong dependence of algorithmic performance on hardware topology and noise characteristics. For example, 10-qubit GHZ states achieved fidelities above 0.8 on trapped-ion hardware, while superconducting platforms dropped below 0.15 due to routing overhead and accumulated two-qubit gate errors. These findings highlight the importance of hardware-aware algorithm selection and provide practical guidance for benchmarking in the NISQ era.
}


\maketitle

\section{Introduction}\label{sec1}

Quantum computing is undergoing rapid progress, evolving from a theoretical concept into a practical computational paradigm. By exploiting quantum mechanical phenomena such as superposition and entanglement~\cite{b1,b2}, quantum computers can outperform classical systems for specific problem classes~\cite{b3}.

Unlike classical computing, which has converged on the von Neumann architecture~\cite{b5}, quantum computing is characterized by diverse physical implementations~\cite{b4}. Leading architectures include trapped-ion and superconducting platforms~\cite{b6,b7}, each offering distinct trade-offs between connectivity, gate speed, and noise characteristics~\cite{b8,b9,b10}. As a result, the same quantum algorithm may yield significantly different outcomes when executed on different devices~\cite{b11}.

Understanding how architectural characteristics translate into algorithm-level performance is therefore essential for both near-term applications and long-term system design~\cite{b12}. While prior studies have benchmarked individual devices or algorithms~\cite{b13,b14,b15,b16,b17,b18}, systematic cross-platform comparisons across modern hardware remain limited.

This paper addresses this gap by presenting a multi-algorithm benchmarking study across state-of-the-art trapped-ion and superconducting quantum processors, with a simulator baseline. We analyze how hardware architecture influences fidelity, circuit depth, success probability, and optimization performance.

\section{Methods and Materials}\label{sec2}

\subsection{Literature Review}\label{subsec1}

The field of quantum computing has progressed rapidly, transitioning from purely theoretical constructs to the availability of small-scale, programmable hardware prototypes. These Noisy Intermediate-Scale Quantum (NISQ) devices, while not yet capable of fault-tolerant computation, offer unprecedented opportunities to execute quantum algorithms and to explore the practical challenges of building scalable quantum machines. A central issue in this era is understanding how to best utilize these emerging technologies, which are built on diverse physical principles and exhibit vastly different architectural characteristics \cite{b16}. This section situates our work within the context of existing hardware platforms and the comparative studies that have sought to benchmark their performance, highlighting the specific architectural trade-offs that influence algorithmic success.

The stark architectural differences between trapped-ion and superconducting platforms have motivated numerous comparative studies aimed at understanding their relative performance. Early work by Linke et al. \cite{b14} directly compared a 5-qubit fully connected ion-trap system with a 5-qubit superconducting system from IBM featuring a star-shaped topology. Running identical algorithms on both, they found that performance was strongly correlated with how well the algorithm’s connectivity requirements matched the hardware’s topology. For example, the hidden-shift algorithm, which required entangling gates between non-adjacent qubits on the IBM device, achieved only a 35 \% success rate, compared to 77 \% on the fully connected ion-trap processor. This study was among the first to experimentally confirm that qubit connectivity is a critical determinant of algorithmic performance and to emphasize the importance of hardware-algorithm co-design.

Subsequent studies expanded on this foundation by employing a variety of benchmarks to probe different performance aspects. Blinov et al. \cite{b13} compared cloud-based systems from IonQ, IBM, and Rigetti using the Bernstein–Vazirani algorithm, demonstrating that the performance of superconducting systems degraded sharply as the number of required CNOT gates exceeded the hardware’s native connectivity, necessitating additional noisy SWAP operations. In contrast, the fully connected IonQ system exhibited only a modest, linear decrease in performance consistent with simple gate-error accumulation. Similarly, Schwaller et al. \cite{b19} implemented Quantum Nondemolition (QND) measurements and observed that IonQ hardware maintained constant state fidelity throughout the QND process, outperforming IBM Q systems—a result attributed to lower gate errors and full qubit connectivity.

More extensive, application-level benchmarking has further reinforced these findings. A large-scale study by Montañez-Barrera et al. \cite{b17} employed a variant of the Quantum Approximate Optimization Algorithm (LR-QAOA) to benchmark 24 processors from six vendors. Their results showed that fully connected systems like those from Quantinuum and IonQ could execute the largest certified instances of fully connected problems, avoiding the substantial SWAP overhead incurred by fixed-layout superconducting devices. The authors also found that component-level metrics such as Error-Per-Layered-Gate (EPLG) did not consistently predict algorithmic performance at scale, underscoring the need for system-level, application-oriented benchmarks.
Complementary work by Zhu et al. \cite{b15} developed cross-platform validation methods using randomized measurements across nine quantum computers from IonQ, UMD, and IBM. They discovered a phenomenon of “intra-technology similarity,” where quantum states generated on devices using the same underlying technology (e.g., two IBM systems) were more similar to each other than to those from different platforms, revealing technology-specific noise signatures.

The existing literature provides a strong foundation for understanding the architectural trade-offs between leading quantum computing platforms. It is well established that the all-to-all connectivity and long coherence times of trapped-ion systems offer a distinct advantage for algorithms requiring complex qubit interactions, while superconducting systems provide faster gate speeds at the cost of limited connectivity and higher error rates. Numerous studies have benchmarked systems from IBM, Rigetti, and IonQ, often concluding that trapped-ion architectures exhibit superior performance on specific algorithmic tasks because of these architectural strengths.

However, a research gap remains in the systematic, head-to-head comparison of algorithm performance across the latest generation of hardware, including IonQ’s Aria-1 and Forte-1, IQM’s Garnet, and Rigetti’s Ankaa-3. Much of the prior comparative work has focused on earlier IBM devices or on single-algorithm studies such as LR-QAOA. Furthermore, several investigations relying on vendor-provided noisy simulators have shown that such models fail to reproduce real device behavior—particularly correlated errors arising from SWAP operations on superconducting chips—highlighting the need for direct empirical validation.
Our study addresses this gap by performing a multi-algorithm, cross-platform benchmarking campaign on these state-of-the-art devices, providing fresh insights into their capabilities and the practical impact of their respective architectural designs.

\subsection{Materials}\label{subsec2}
\subsubsection{IonQ Forte and Aria}
IonQ processors are based on trapped-ion technology using $^{171}\mathrm{Yb}^{+}$ ions~\cite{b20}. Native all-to-all connectivity enables direct entangling operations between arbitrary qubit pairs, reducing routing overhead and circuit depth.

\subsubsection{IQM Garnet}\label{subsubsec2}
IQM Garnet is a 20-qubit superconducting transmon processor with nearest-neighbor connectivity~\cite{b21}. Two-qubit interactions require routing operations that increase circuit depth for non-local gates.

\subsubsection{Rigetti Ankaa-3}\label{subsubsec3}
Rigetti Ankaa-3 is an 84-qubit superconducting processor featuring tunable couplers and a CZ-based native gate set~\cite{b22}. While scalable, its two-dimensional layout imposes connectivity constraints.

\subsubsection{State Vector Simulator (SV1)}\label{subsubsec4}
Amazon Braket SV1 provides ideal, noise-free simulation of quantum circuits up to 34 qubits~\cite{b23}, serving as a reference baseline.

\subsection{Algorithm Benchmarks}

\subsubsection{The Bell State}\label{subsubsec5}
To benchmark the performance of the quantum devices under study, we first implemented the Bell state algorithm, a standard test for bipartite entanglement. The Bell state serves as a minimal circuit that is highly sensitive to gate errors, decoherence, and readout noise, making it an effective probe of device-level fidelity \cite{b24,b26}.

The circuit begins with two qubits initialized in the $\ket{00}$ state. A Hadamard gate is applied to the first qubit, placing it into a superposition. Subsequently, a controlled-NOT (CNOT) gate is applied with the first qubit as the control and the second qubit as the target, resulting in the entangled Bell state:

\begin{figure}[h]
\centering
\includegraphics[width=0.9\linewidth]{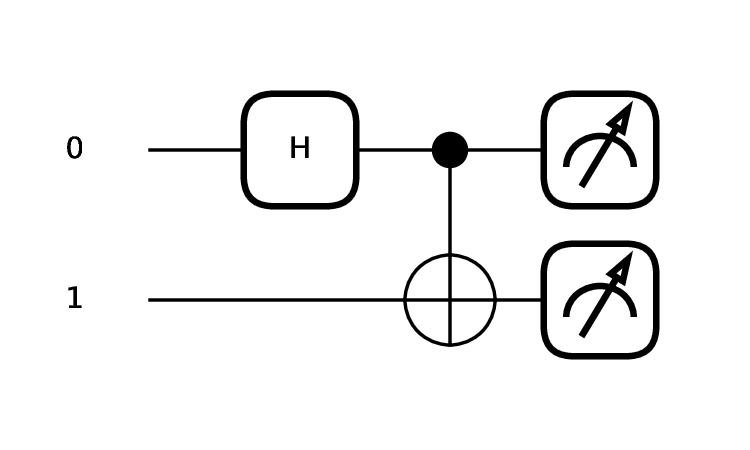}
\caption{Quantum circuit used to prepare the Bell state $|\Phi^+\rangle$.}
\label{fig:bell_circuit}
\end{figure}

\begin{equation}
|\Phi^+\rangle = \frac{1}{\sqrt{2}}\left( |00\rangle + |11\rangle \right)
\end{equation}
This circuit was implemented on all selected devices — IonQ Aria-1, IonQ Forte-1, IQM Garnet, Rigetti Ankaa-3, and the Amazon SV1 simulator.

To quantify nonlocal correlations, the CHSH parameter
\begin{equation}
S = E(a,b) + E(a,b') + E(a',b) - E(a',b')
\end{equation}
was evaluated. Values exceeding the classical bound $S \le 2$ indicate entanglement~\cite{b25}. Experimental values slightly exceeding the Tsirelson bound are attributed to finite sampling statistics and readout bias.

\subsubsection{GHZ state}\label{subsubsec6}
To extend our benchmarking beyond bipartite entanglement, we implemented the Greenberger–Horne–Zeilinger (GHZ) algorithm, which creates maximally entangled states of three or more qubits. GHZ states are more sensitive to noise and decoherence than Bell states, making them a standard benchmark for assessing scalability of quantum hardware \cite{b27}.

The GHZ circuit starts with $n$ qubits initialized in the $|0\rangle^{\otimes n}$ state. A Hadamard gate is applied to the first qubit to create superposition. Then, a sequence of controlled-NOT (CNOT) gates is applied from the first qubit to each of the remaining qubits. For three qubits, the resulting state is:

\begin{figure}[h]
\centering
\includegraphics[width=0.9\linewidth]{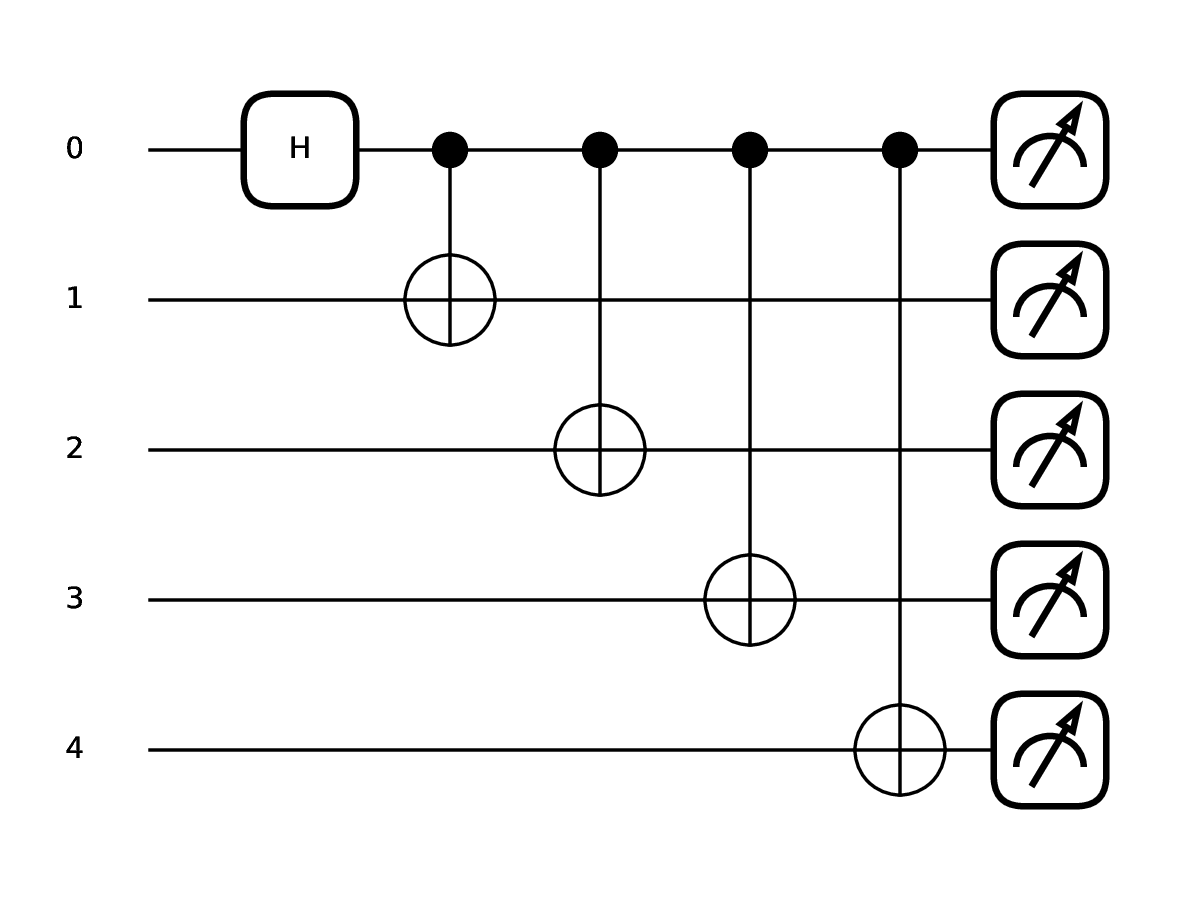}
\caption{Circuit diagram for preparing an $n$-qubit GHZ state.}
\label{fig:ghz_circuit}
\end{figure}

\begin{equation}
    |\text{GHZ}_3\rangle = \frac{1}{\sqrt{2}}\left( |000\rangle + |111\rangle \right)
\end{equation}
This circuit generalizes to $n$ qubits as:
\begin{equation}
    |\text{GHZ}_n\rangle = \frac{1}{\sqrt{2}}\left( |0\rangle^{\otimes n} + |1\rangle^{\otimes n} \right)
\end{equation}

We evaluated the following performance indicators: \textit{Fidelity} with respect to the ideal GHZ state:
\begin{equation}
    F_{\text{GHZ}} = \langle \text{GHZ}_n | \rho_{\text{exp}} | \text{GHZ}_n \rangle
\end{equation}
where $\rho_{\text{exp}}$ is the experimental density matrix obtained via tomography \cite{b1}.

\textit{Coherence} across multiple qubits was probed by applying local phase rotations and measuring parity oscillations, a standard technique to confirm GHZ coherence \cite{b28}.

Low fidelities observed on superconducting devices at larger qubit counts arise from accumulated two-qubit gate errors and SWAP operations required by limited connectivity.

\subsubsection{Quantum Fourier Transform (QFT)}\label{subsubsec7} 
Fundamental subroutine in many quantum algorithms, including Shor’s factoring algorithm and quantum phase estimation \cite{b1,b29}. In this study, we implemented QFT circuits to benchmark circuit depth, gate decomposition, and device noise when scaling beyond small entangled states.
For $n$ qubits, the QFT is defined as the linear transformation:
\begin{equation}
    |x\rangle \;\mapsto\; \frac{1}{\sqrt{2^n}} \sum_{y=0}^{2^n - 1} e^{2\pi i xy / 2^n} |y\rangle
\end{equation}
where $\ket{x}$ is a computational basis state.\\
The QFT circuit consists of:\\
1. Hadamard gate on each qubit,\\
2. Controlled-phase rotations $R_k = \operatorname{diag}\left(1, e^{2\pi i / 2^k}\right)$ applied between qubits,\\
3. A final reversal (SWAP gates) to invert qubit order.

\begin{figure}[h]
\centering
\includegraphics[width=0.9\linewidth]{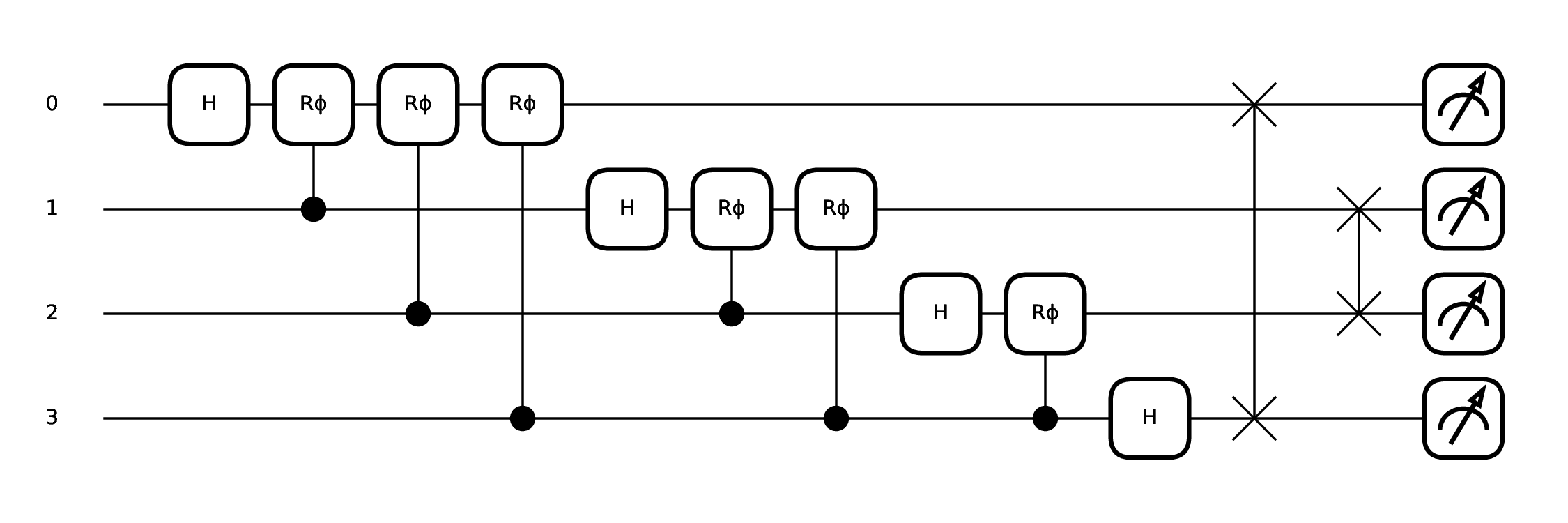}
\caption{General structure of the $n$-qubit Quantum Fourier Transform (QFT).}
\label{fig:qft_circuit}
\end{figure}

We measured logical circuit depth, number of single- and two-qubit gates, and the frequency of controlled rotations. These metrics indicate scalability challenges and hardware efficiency \cite{b31}.

For larger qubit counts, approximate QFT can be implemented by discarding small-angle controlled rotations. The error is quantified as:
\begin{equation}
    \epsilon = \| U_{\text{QFT}} - U_{\text{approx}} \|
\end{equation}
where $U_{\text{QFT}}$ is the exact unitary and $U_{\text{approx}}$ is the truncated version \cite{b30}.

To benchmark hardware performance, we implemented QFT followed by its inverse (IQFT). Ideally, the operation should return the input state. The fidelity is:
\begin{equation}
    F = \langle \psi | U_{\text{IQFT}} \, U_{\text{QFT}} | \psi \rangle
\end{equation}
where $\ket{\psi}$ is the initial state. This provides a hardware-dependent measure of QFT reliability.

\subsubsection{Grover’s Search}\label{subsubsec8}
The algorithm provides a quadratic speedup for unstructured search problems, reducing the query complexity from $O(N)$ to $O(\sqrt{N})$ \cite{b32}. We used Grover’s search as a benchmark for iterative amplitude amplification, circuit depth scaling, and hardware reliability when running multiple iterations of oracle and diffusion operators.

Given a search space of size $N=2^n$ with a marked state $\ket{\omega}$ Grover's algorithm applies the following sequence:

1. Initialization: Apply Hadamard gates to all $n$ qubits to create a uniform superposition:
\begin{equation}
    |\psi\rangle = \frac{1}{\sqrt{N}} \sum_{x=0}^{N-1} |x\rangle
\end{equation}
2. Oracle $O_\omega$: Marks the solution by applying a phase flip:

\begin{equation}
    O_w |x\rangle =
\begin{cases}
- |x\rangle, & \text{if } x = w, \\
\ \ |x\rangle, & \text{otherwise.}
\end{cases}
\end{equation}

3. Diffusion operator $D$: Inverts amplitudes about the mean:
\begin{equation}
    D = 2|\psi\rangle\langle\psi| - I
\end{equation}

4. Iteration: Repeat $k$ times, where the optimal number is approximately:
\begin{equation}
    k^* \approx \left\lfloor \frac{\pi}{4} \sqrt{\frac{N}{M}} \right\rfloor
\end{equation}
with $M$ being the number of marked states.

\begin{figure}[h]
\centering
\includegraphics[width=0.9\linewidth]{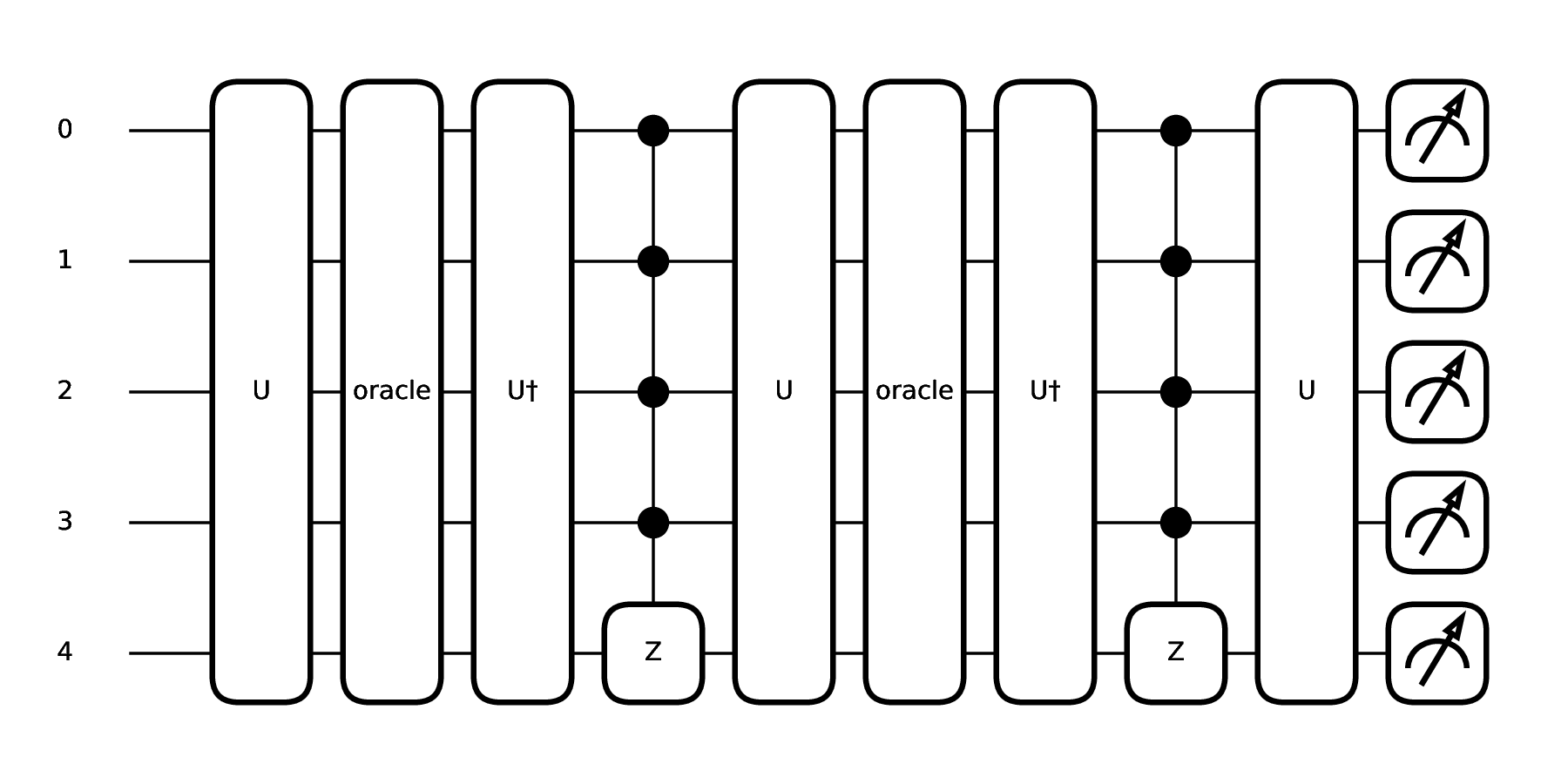}
\caption{Grover's algorithm iteration: oracle followed by diffusion operator.}
\label{fig:grover_circuit}
\end{figure}

Measured as the probability of measuring the marked state after $k$ iterations:
\begin{equation}
    p_{\text{success}}(k) = \left| \langle w | G^k | \psi \rangle \right|^2
\end{equation}
where $G = DO_\omega$ is the Grover iterate.

We compared how the \textit{success probability} and \textit{circuit depth} scale with the number of qubits across different devices \cite{b33}.

To evaluate hardware constraints—particularly under limited connectivity—we extracted the \textit{logical circuit depth, number of multi-qubit gates}, and \textit{total gate} counts \cite{b1}.

Finally, we compared the observed iteration count that maximized $p_{\textit{success}}$ with the theoretical $k^*$, quantifying deviations arising from noise.

\subsubsection{Quantum Approximate Optimization Algorithm (QAOA)}\label{subsubsec9}
Variational hybrid quantum-classical algorithm designed for solving combinatorial optimization problems \cite{b34,b35}. In this work, we implemented QAOA on the Minimum Vertex Cover problem to benchmark variational performance across different hardware platforms.

QAOA prepares a parameterized quantum state by alternating between a problem Hamiltonian and a mixing Hamiltonian:

1. Problem Hamiltonian: Encodes the cost function $C(z)$ into a diagonal Hamiltonian $H_C$ such that eigenvalues correspond to objective function values. 
\begin{equation}
    H_C |z\rangle = C(z) |z\rangle
\end{equation}

2. Mixing Hamiltonian: Promotes exploration of the solution space using:
\begin{equation}
    H_M = i \sum_i X_i
\end{equation}
where $X_i$ are Pauli-X operators.

3. Parameterized Circuit:
For depth $p$, the QAOA state is: 
\begin{equation}
    |\gamma, \beta\rangle = \prod_{k=1}^{p} \left( e^{-i \beta_k H_M} e^{-i \gamma_k H_C} \right) |+\rangle^{\otimes n}
\end{equation}

4. Optimization Loop:
Classical optimizers (COBYLA, SPSA, etc.) are used to select angles $(\gamma, \beta)$ that minimize the expected cost:
\begin{equation}
    \langle H_C \rangle = \langle \gamma, \beta | H_C | \gamma, \beta \rangle
\end{equation}

\begin{figure}[h]
\centering
\includegraphics[width=0.45\columnwidth]{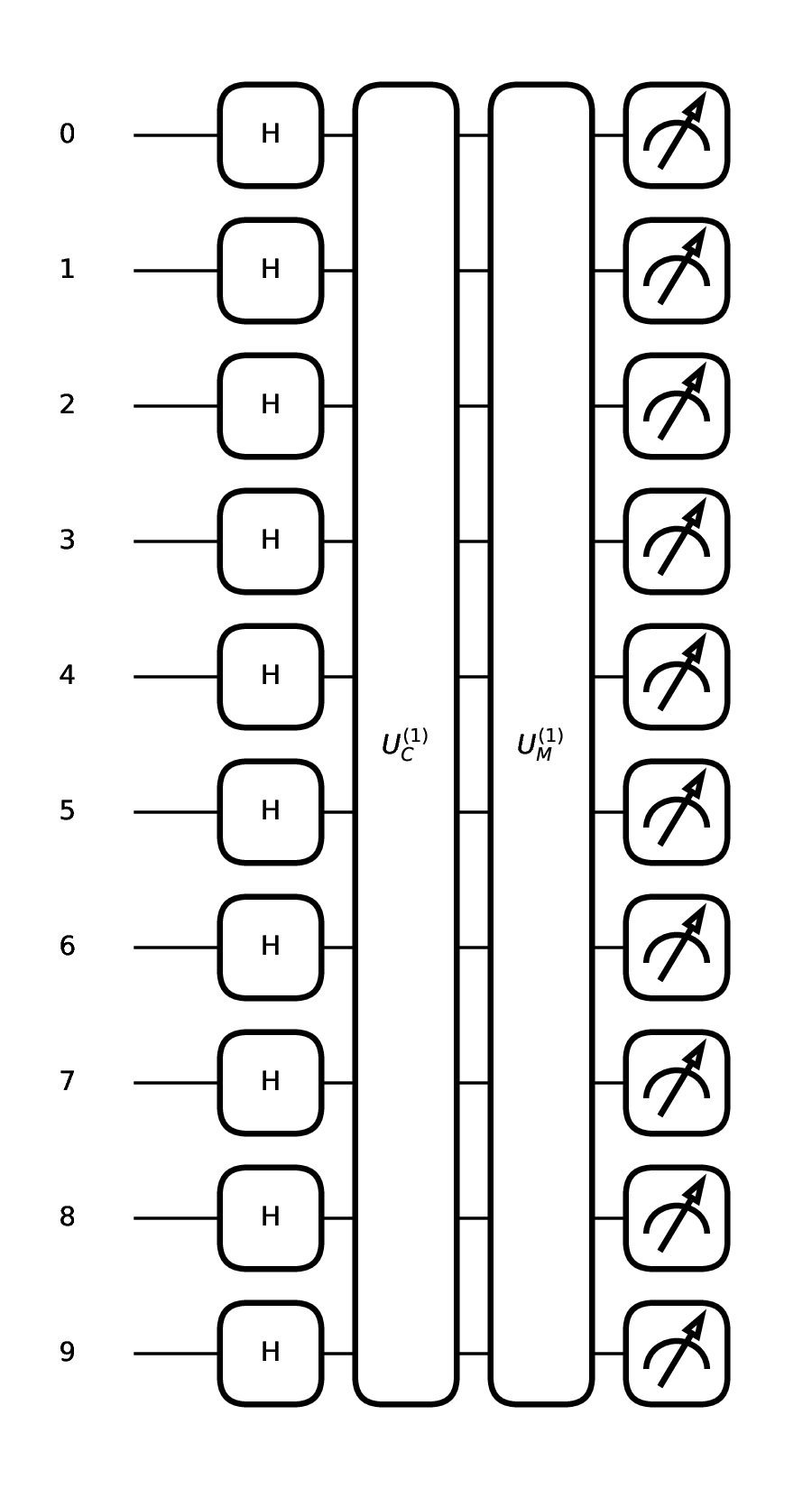}
\caption{Structure of a depth-$p$ Quantum Approximate Optimization Algorithm.}
\label{fig:qaoa_circuit}
\end{figure}

To evaluate algorithmic performance and hardware behavior, we considered several key metrics \cite{b2}. 
The \textit{approximation ratio (AR)}, defined as the ratio between the optimal cost and the observed expectation value of the cost Hamiltonian,
\begin{equation}
    AR = \frac{C_{\text{opt}}}{\langle H_C \rangle},
\end{equation}
quantifies how closely the algorithm approaches the optimal solution \cite{b37}. 
The \textit{success probability} measures the likelihood of sampling the bitstring corresponding to the optimal solution, reflecting the circuit’s reliability and consistency across repeated executions \cite{b2, b33}. 

The \textit{feasibility rate} represents the proportion of experimentally obtained bitstrings that satisfy the constraints of the optimization problem, thereby linking hardware precision with the validity of the generated solutions \cite{b2, b38}. 

The \textit{Hamming distance} between measured and optimal bitstrings provides a measure of similarity; its mean value indicates how close the sampled solutions are to the ideal configuration, while its variance captures the influence of noise, decoherence, and sampling errors \cite{b31,b33}. 

Finally, \textit{training convergence} characterizes the behavior of the classical optimization loop within QAOA. Faster and more stable convergence indicates efficient parameter optimization and a balanced interplay between the quantum and classical components \cite{b2}. 
In addition, \textit{gate and depth analysis}---including logical circuit depth, multi-qubit gate count, and total gate operations---was used to assess device constraints, especially under limited connectivity \cite{b2,b39}. 

\subsection{Methodology}\label{subsec3}

All algorithms in this study were implemented and executed using the Amazon Braket quantum computing service to ensure a consistent and reproducible experimental environment. The selected devices represent a variety of quantum hardware architectures, allowing for comparative evaluation of their operational fidelity, connectivity, and noise characteristics. The experiments were conducted on the following platforms:

\begin{itemize}
\item \textbf{SV1 (simulator)} – a state-vector simulator providing a noise-free reference baseline.
\item \textbf{IonQ Aria-1 and IonQ Forte-1 (QPU)} – trapped-ion architectures featuring full qubit connectivity and long coherence times.
\item \textbf{IQM Garnet (QPU, 20 qubits)} – a superconducting transmon-based system with nearest-neighbor connectivity.
\item \textbf{Rigetti Ankaa-3 (QPU)} – a superconducting processor with tunable couplers and intermediate-scale qubit capacity.
\end{itemize}

All quantum circuits were constructed using the \textit{PennyLane} framework~\cite{b39} and executed via the \textit{Amazon Braket SDK}~\cite{b40}, with results automatically stored for post-processing. For each circuit configuration, \textit{100 shots} were performed to obtain statistically meaningful measurement outcomes. Where applicable, the same compilation settings, qubit mapping strategy, and readout structure were applied across devices to ensure comparability. Although PennyLane provides a unified interface, backend-specific compilation and native gate decompositions may still introduce unavoidable device-dependent variations.

For the \textit{entanglement-based benchmarks} (Bell and GHZ states), the experiments were primarily designed to probe hardware fidelity and noise sensitivity. Each Bell circuit (Fig.~\ref{fig:bell_circuit}) used two qubits prepared in the maximally entangled Bell state and measured in the appropriate CHSH bases to extract the Bell parameter $S$. GHZ circuits (Fig.~\ref{fig:ghz_circuit}) extended this to multipartite entanglement using $n = 6$ and $n = 10$ qubits. State fidelities and density-matrix reconstructions were used to quantify multi-qubit coherence, while CHSH inequality violations characterized bipartite entanglement quality across architectures. In the results, we focus on representative GHZ configurations at $n = 6$ and $n = 10$ qubits to highlight scaling behavior.

The \textit{Quantum Fourier Transform (QFT)} benchmark (Fig.~\ref{fig:qft_circuit}) was implemented as a QFT–IQFT round-trip circuit to evaluate phase-coherent multi-qubit operations. A QFT layer was followed immediately by its inverse (IQFT), ideally returning the system to its initial computational basis state. Round-trip fidelity was defined as the probability of correctly recovering the input bitstring after the QFT–IQFT sequence, thereby capturing the cumulative effect of phase-rotation accuracy, two-qubit gate fidelity, and routing overhead. We report results for representative register sizes of $n = 6$ and $n = 10$ qubits.

For \textit{Grover’s search}, circuits (Fig.~\ref{fig:grover_circuit}) were constructed for small unstructured search spaces with a single marked element. The oracle and diffusion operators were applied iteratively, and the success probability $P_{\mathrm{success}}$ was evaluated as a function of the iteration number $k$. For each device and qubit configuration, three circuits corresponding to $k-1$, $k$, and $k+1$ iterations were executed to probe the sensitivity of amplitude amplification to over- and under-rotation. The reported results focus on four- and six-qubit search spaces.

The \textit{Quantum Approximate Optimization Algorithm (QAOA)} was applied to the Minimum Vertex Cover problem on graph instances of 10 nodes, including $Path_{10}$ (sparse), $BA_{10,2}$ (moderate connectivity), and $K_{5,5}$ (dense). Circuits (Fig.~\ref{fig:qaoa_circuit}) were executed with depth $p = 1$ in a hybrid quantum–classical loop, where variational parameters were optimized using a classical optimizer based on repeated circuit evaluations. From the resulting bitstring samples, we computed the approximation ratio $r = C_{\mathrm{quantum}} / C_{\mathrm{opt}}$, feasibility rate (fraction of valid vertex covers), success probability of sampling an optimal solution, and the average Hamming distance between measured and optimal bitstrings. Logical circuit depth and gate composition were extracted from compiled circuits to analyze hardware-dependent overhead.

All collected experimental data were processed using Python-based analysis pipelines built with \textit{NumPy}, \textit{Pandas}, and \textit{Matplotlib}. Fidelity, success probabilities, and approximation ratios were computed for each device, and error bars were estimated from the empirical variance across repeated runs or equivalent configurations. This workflow enabled direct quantitative comparison between devices and algorithms, highlighting the trade-offs between circuit complexity, noise resilience, and architectural scalability.

\section{Results}\label{sec3}

This section presents the experimental results obtained for all five quantum algorithms implemented on multiple hardware backends using Amazon Braket~\cite{b40}. Each benchmark was executed on the SV1 simulator and a subset of trapped-ion (IonQ Aria-1 or IonQ Forte-1) and superconducting (IQM Garnet, Rigetti Ankaa-3) devices, enabling a comparative analysis of how architectural differences affect performance. The results are evaluated through several quantitative metrics, including state fidelity, CHSH violation, success probability, approximation ratio, and circuit depth. Together, these outcomes provide a comprehensive view of the relationship between algorithmic complexity and device-level noise characteristics.

\begin{table}[h]
\caption{Summary of benchmark algorithms, qubit sizes, devices, and primary evaluation metrics}
\label{tab:quantum_algorithms}
\centering
\begin{tabular*}{\textwidth}{@{\extracolsep\fill}lcccc}
\toprule
Algorithm & Qubits & Devices Tested & Primary Metric & Purpose \\
\midrule
Bell   & 2     & IonQ, IQM, Rigetti, SV1 & Fidelity, CHSH $S$        & Entanglement benchmark \\
GHZ    & 6, 10 & IonQ, IQM, Rigetti, SV1 & Fidelity, coherence       & Multipartite entanglement \\
QFT    & 6, 10 & IonQ, IQM, Rigetti, SV1 & Round-trip fidelity       & Phase-coherence benchmark \\
Grover & 4, 6  & IonQ, IQM, Rigetti, SV1 & Success probability       & Algorithmic scaling \\
QAOA   & 10    & IonQ, IQM, Rigetti, SV1 & Approximation ratio       & Optimization benchmark \\
\bottomrule
\end{tabular*}
\end{table}

\subsection{Bell State Results}\label{subsec4}

As summarized in Table~\ref{tab:quantum_algorithms}, the Bell benchmark uses two qubits to probe bipartite entanglement via the CHSH parameter $S$. The ideal Bell state
\[
\lvert \Phi^{+} \rangle = \frac{1}{\sqrt{2}} \left( \lvert 00 \rangle + \lvert 11 \rangle \right)
\]
was prepared and measured in the CHSH bases on both simulated and physical quantum devices. The Clauser–Horne–Shimony–Holt (CHSH) parameter $S$ was computed to quantify the violation of Bell’s inequality, where classical correlations satisfy $S \leq 2$ and the quantum mechanical maximum is $S_{\text{ideal}} = 2\sqrt{2} \approx 2.828$.

\begin{table}[h]
\caption{Experimental CHSH values $S_\text{exp}$ measured on each device, compared against the ideal Bell value $S_\text{ideal}=2.828$}
\label{tab:chsh_results}
\centering
\scriptsize
\begin{tabular*}{\textwidth}{@{\extracolsep\fill}l l c c c}
\toprule
Device & Type & $S_\text{exp}$ & $\pm$Err. & $S_\text{ideal}$ \\
\midrule
SV1 (Simulator)        & State vector                    & 2.7 & 0.15 & 2.828 \\
IonQ Aria-1 (QPU)      & Trapped ion (all-to-all)         & 2.9 & 0.14 & 2.828 \\
Rigetti Ankaa-3 (QPU)  & Superconducting (NN)             & 2.2 & 0.16 & 2.828 \\
IQM Garnet (QPU, 20q)  & Superconducting (NN)             & 2.5 & 0.16 & 2.828 \\
\bottomrule
\end{tabular*}
\end{table}

\begin{figure}[htbp]
\centering
\includegraphics[width=\columnwidth]{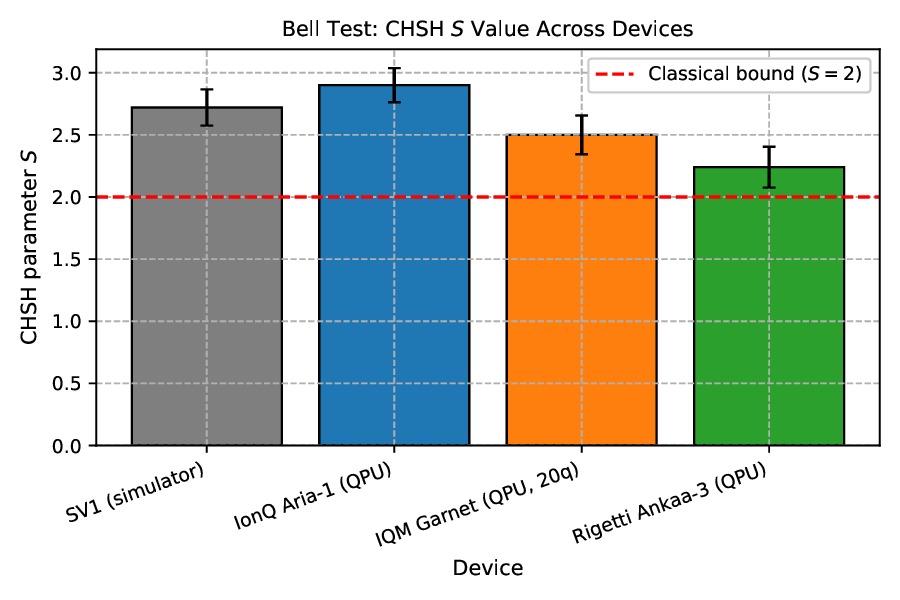}
\caption{CHSH parameter $S$ for the Bell state experiment on each device. The dashed line marks the classical bound $S = 2$.}
\label{fig:chsh_s_values}
\end{figure}

Table~\ref{tab:chsh_results} summarizes the experimental CHSH values across all tested platforms, while Fig.~\ref{fig:chsh_s_values} provides a visual comparison. The SV1 simulator produced $S_{\text{exp}} \approx 2.7 \pm 0.15$, closely matching the ideal quantum prediction and confirming correct circuit implementation in the absence of noise. Among the physical quantum processors, IonQ Aria-1 achieved the highest experimental value $S_{\text{exp}} \approx 2.9 \pm 0.14$, slightly exceeding the theoretical bound within statistical uncertainty and reflecting strong entanglement generation enabled by its all-to-all ion-trap connectivity and long coherence times.

In contrast, the superconducting devices Rigetti Ankaa-3 and IQM Garnet demonstrated lower CHSH values (approximately $2.2 \pm 0.16$ and $2.5 \pm 0.16$, respectively), consistent with their nearest-neighbor connectivity and higher two-qubit gate error rates. Despite these reductions, all devices exhibited $S > 2$, thereby confirming the successful observation of quantum entanglement across all architectures and highlighting the impact of hardware connectivity and noise on the preservation of non-classical correlations.

\subsection{GHZ State Results}\label{subsec5}

As indicated in Table~\ref{tab:quantum_algorithms}, the GHZ benchmark extends entanglement to $n = 6$ and $n = 10$ qubits to study multipartite coherence. On the SV1 simulator, GHZ states exhibited fidelities close to unity, confirming the absence of decoherence and ideal gate behavior. When executed on quantum hardware, the fidelity systematically decreased with the number of qubits, illustrating the cumulative effect of decoherence, gate errors, and imperfect calibration.

\begin{figure}[htbp]
\centering
\includegraphics[width=\columnwidth]{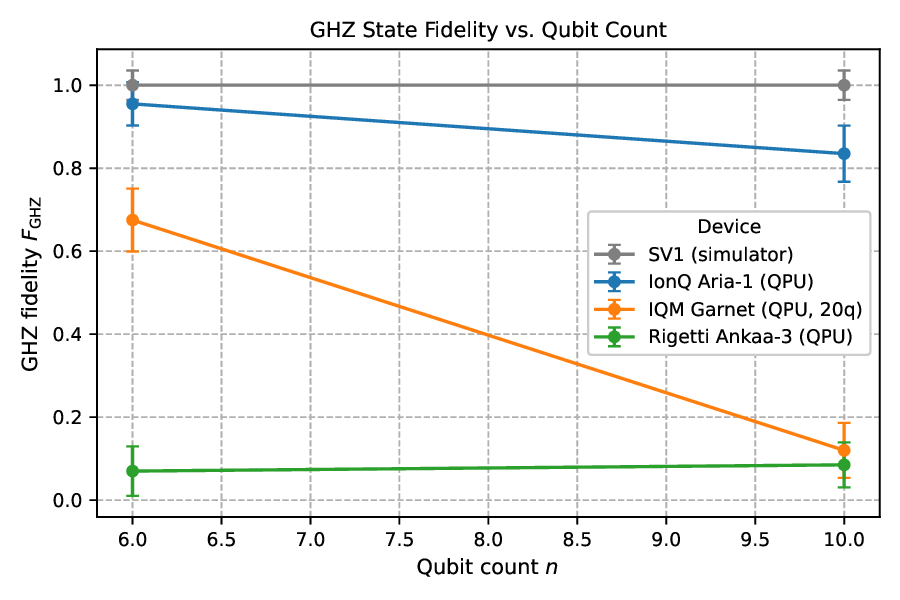}
\caption{GHZ state fidelities $F_{\text{GHZ}}$ as a function of qubit count $n$ across the tested devices.}
\label{fig:ghz_fidelity_vs_qubits}
\end{figure}

Table~\ref{tab:ghz_fidelity_results} reports representative fidelities for $n = 6$ and $n = 10$ qubits, while Fig.~\ref{fig:ghz_fidelity_vs_qubits} shows how $F_{\text{GHZ}}$ scales with qubit count across devices. The trapped-ion IonQ Aria-1 maintained the highest GHZ fidelities, achieving $F_{\text{exp}} \approx 0.96$ at $n = 6$ and $F_{\text{exp}} \approx 0.84$ at $n = 10$. IQM Garnet and Rigetti Ankaa-3 exhibited significantly stronger degradation, dropping to $F_{\text{exp}} \approx 0.12$ and $0.09$ at $n = 10$, respectively. This behavior is particularly pronounced on superconducting architectures, where limited qubit connectivity necessitates additional SWAP operations that compound error accumulation.

These results demonstrate that all tested devices can generate multipartite entanglement, yet only the trapped-ion architecture sustains high-quality coherence as the qubit count increases. The observed fidelity degradation trend is consistent with prior reports on Noisy Intermediate-Scale Quantum (NISQ) systems~\cite{b1}.  

\begin{table}[h]
\caption{Experimental GHZ fidelities $F_{\text{exp}}$ for $n=6$ and $n=10$ qubits, with ideal noiseless values $F_{\text{ideal}}$ shown for reference}
\label{tab:ghz_fidelity_results}
\centering
\scriptsize
\begin{tabular*}{\textwidth}{@{\extracolsep\fill}l c c c c c}
\toprule
Device & Qubits ($n$) & $F_{\text{exp}}$ & $\pm$Err. & $F_{\text{ideal}}$ & Shots/setting \\
\midrule
SV1 (Simulator)          & 6  & 1.000 & 0.035 & 1.000 & 100 \\
SV1 (Simulator)          & 10 & 1.000 & 0.035 & 1.000 & 100 \\
IonQ Aria-1 (QPU)        & 6  & 0.955 & 0.052 & 1.000 & 100 \\
IonQ Aria-1 (QPU)        & 10 & 0.835 & 0.068 & 1.000 & 100 \\
IQM Garnet (QPU, 20q)    & 6  & 0.675 & 0.076 & 1.000 & 100 \\
IQM Garnet (QPU, 20q)    & 10 & 0.12  & 0.066 & 1.000 & 100 \\
Rigetti Ankaa-3 (QPU)    & 6  & 0.07  & 0.06  & 1.000 & 100 \\
Rigetti Ankaa-3 (QPU)    & 10 & 0.09  & 0.054 & 1.000 & 100 \\
\bottomrule
\end{tabular*}
\end{table}

\subsection{Quantum Fourier Transform (QFT) Results}\label{subsec6}

\begin{figure}[htbp]
\centering
\includegraphics[width=\columnwidth]{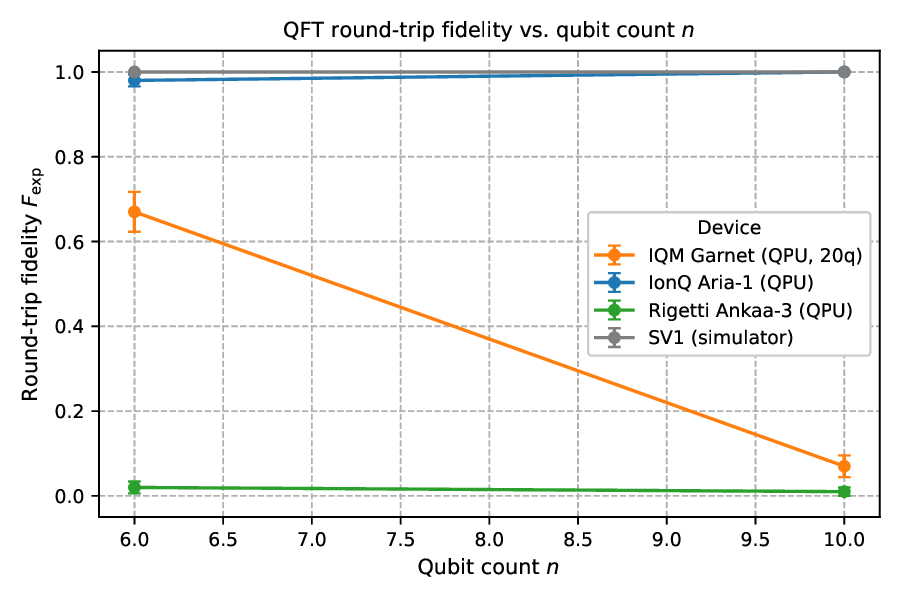}
\caption{QFT round-trip fidelities $F_{\mathrm{exp}}$ as a function of qubit count $n$. Error bars denote statistical uncertainties, and solid lines indicate individual devices.}
\label{fig:qft_fexp_vs_n}
\end{figure}

\begin{figure}[htbp]
\centering
\includegraphics[width=\columnwidth]{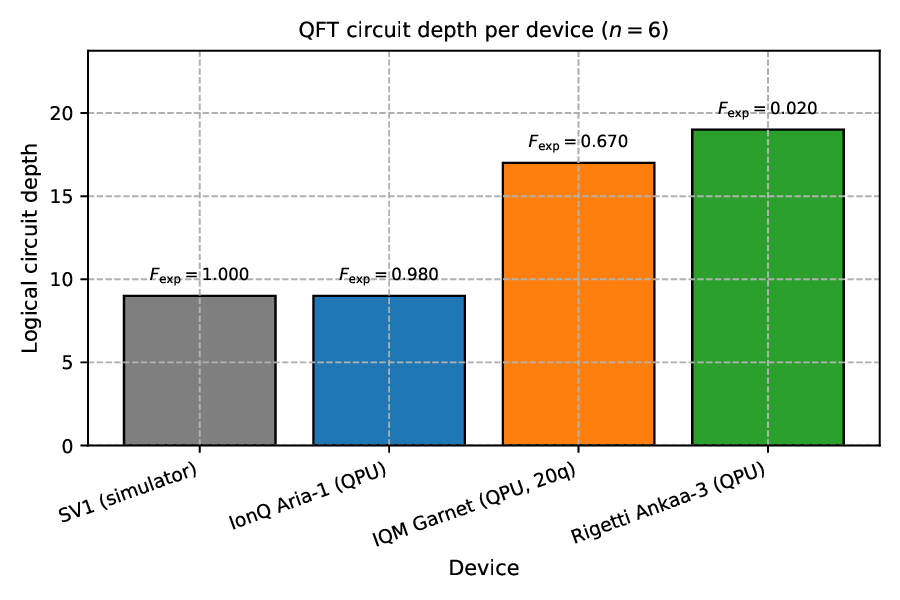}
\caption{Estimated logical circuit depth per device for a QFT round-trip circuit with $n = 6$ qubits. Bars are annotated with the corresponding experimental fidelities $F_{\mathrm{exp}}$.}
\label{fig:qft_depth_per_device}
\end{figure}

The QFT benchmark (Table~\ref{tab:quantum_algorithms}) evaluates round-trip fidelity for $n = 6$ and $n = 10$ qubits using the QFT–IQFT protocol described in the methodology. Table~\ref{tab:qft_roundtrip_fidelity} summarizes the resulting round-trip fidelities and circuit depths for $n = 6$ and $n = 10$ qubits, and Figs.~\ref{fig:qft_fexp_vs_n} and~\ref{fig:qft_depth_per_device} visualize how performance depends on qubit count and architecture.

On the SV1 simulator, the round-trip fidelity remained effectively ideal ($F_{\mathrm{exp}} = 1.00$ within statistical error) for both qubit sizes, confirming the correctness of the compiled QFT and IQFT circuits. On quantum hardware, fidelity decreased markedly with qubit count, reflecting the cumulative impact of gate infidelity and decoherence. The IonQ device achieved the best overall performance, with $F_{\mathrm{exp}} \approx 0.98$ at $n = 6$ and $F_{\mathrm{exp}} \approx 1.00$ at $n = 10$ within the reported uncertainties. IQM Garnet and Rigetti Ankaa-3 exhibited substantially lower fidelities, particularly at $n = 10$, where $F_{\mathrm{exp}}$ fell to $\approx 0.07$ and $\approx 0.01$, respectively.

Circuit depth analysis revealed significant architectural differences. Superconducting devices (Rigetti and IQM) required much deeper circuits due to limited nearest-neighbor connectivity, which increases the number of SWAP operations required for multi-qubit phase rotations. In contrast, the trapped-ion architecture benefited from all-to-all connectivity, resulting in shallower circuits and reduced error accumulation. These results highlight the critical influence of hardware topology on algorithmic scalability and phase-preserving operations.

\begin{table}[h]
\caption{QFT round-trip fidelities $F_{\mathrm{exp}}$ and corresponding logical circuit depths for each device and qubit count. Ideal fidelities $F_{\mathrm{ideal}}=1.000$ provide a noiseless simulation baseline}
\label{tab:qft_roundtrip_fidelity}
\centering
\scriptsize
\begin{tabular*}{\textwidth}{@{\extracolsep\fill}l c c c c c}
\toprule
Device & Qubits ($n$) & $F_{\text{exp}}$ & $\pm$Err. & Circuit depth & $F_{\text{ideal}}$ \\
\midrule
SV1 (Simulator)          & 6  & 1.000 & 0.000 & 9  & 1.000 \\
SV1 (Simulator)          & 10 & 1.000 & 0.000 & 17 & 1.000 \\
IonQ Aria-1 (QPU)        & 6  & 0.980 & 0.014 & 9  & 1.000 \\
IonQ Aria-1 (QPU)        & 10 & 1.000 & 0.000 & 17 & 1.000 \\
IQM Garnet (QPU, 20q)    & 6  & 0.670 & 0.047 & 17 & 1.000 \\
IQM Garnet (QPU, 20q)    & 10 & 0.070 & 0.026 & 47 & 1.000 \\
Rigetti Ankaa-3 (QPU)    & 6  & 0.020 & 0.014 & 19 & 1.000 \\
Rigetti Ankaa-3 (QPU)    & 10 & 0.010 & 0.010 & 51 & 1.000 \\
\bottomrule
\end{tabular*}
\end{table}

\subsection{Grover’s Search Results}

\begin{figure}[htbp]
\centering
\includegraphics[width=\columnwidth]{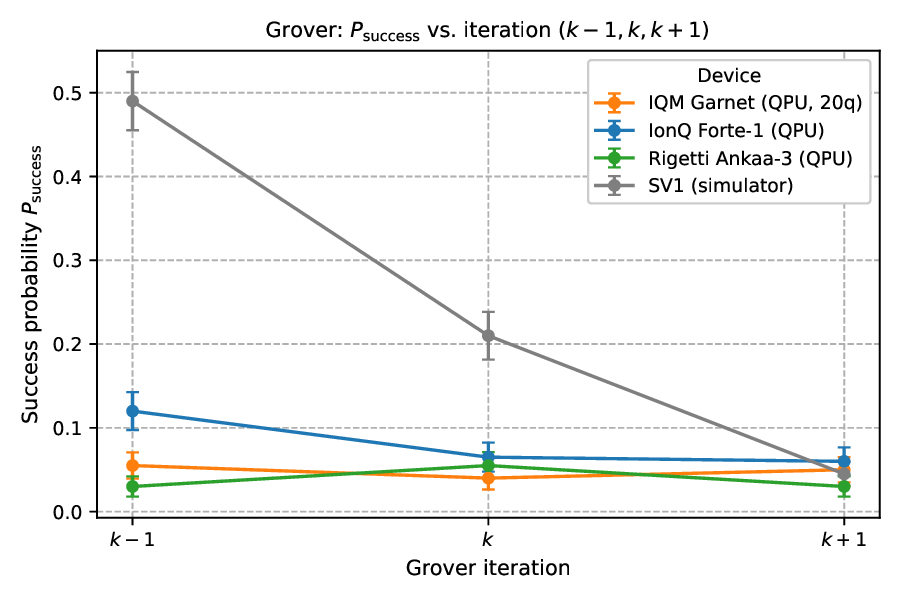}
\caption{Grover search success probabilities $P_{\mathrm{success}}$ as a function of Grover iteration number ($k-1$, $k$, and $k+1$) for each device.}
\label{fig:grover_success_vs_iteration}
\end{figure}

\begin{figure}[htbp]
\centering
\includegraphics[width=\columnwidth]{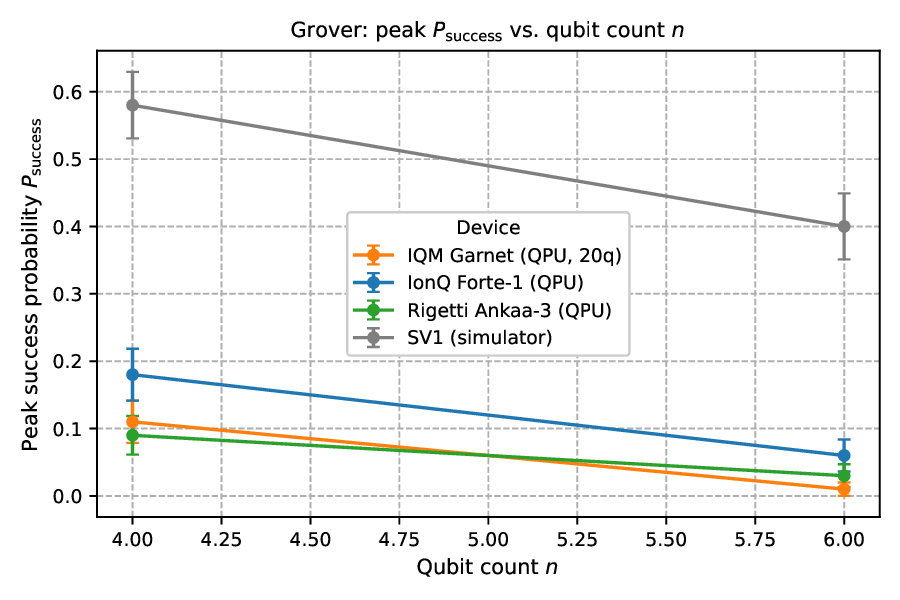}
\caption{Peak experimental success probabilities $P_{\mathrm{success}}$ versus qubit count $n$ for each quantum device.}
\label{fig:grover_peak_success_vs_qubits}
\end{figure}

As listed in Table~\ref{tab:quantum_algorithms}, Grover’s algorithm was benchmarked for $n = 4$ and $n = 6$ qubits by measuring the success probability as a function of the iteration number $k$.
The algorithm was used to benchmark the implementation of iterative amplitude amplification on each device. Table~\ref{tab:grover_success_table} reports the measured success probabilities $P_{\mathrm{success}}$ for four- and six-qubit search spaces at three iteration counts ($k-1$, $k$, $k+1$), Figs.~\ref{fig:grover_success_vs_iteration} and~\ref{fig:grover_peak_success_vs_qubits} summarize the dependence on iteration number and qubit count.

On the simulator, the success probabilities exhibit the expected dependence on the iteration number, with the highest $P_{\mathrm{success}}$ occurring near the optimal iteration. For the instances considered, the peak success probabilities reach approximately $0.58$ for $n = 4$ and $0.40$ for $n = 6$ at $k-1$, with a rapid decline when the number of iterations is increased or decreased. On physical quantum hardware, success probabilities are substantially lower and often close to zero for the larger search spaces, indicating strong sensitivity to gate errors and decoherence. The IonQ Forte-1 device achieves the highest hardware success probabilities, reaching up to $\approx 0.18$ for $n = 4$, whereas IQM Garnet and Rigetti Ankaa-3 typically remain below $\approx 0.10$ across the tested configurations.

\begin{table}[h]
\caption{Grover search benchmarking with $k\!-\!1$, $k$, and $k\!+\!1$ iterations per device and qubit count. Reported are experimental success probabilities $P_{\mathrm{success}}$ with statistical uncertainties}
\label{tab:grover_success_table}
\centering
\scriptsize
\begin{tabular*}{\textwidth}{@{\extracolsep\fill}l c c c c}
\toprule
Device & Qubits ($n$) & Iteration ($k$) & $P_{\mathrm{success}}$ & $\pm$Err. \\
\midrule
IQM Garnet         & 4 & 2 ($k\!-\!1$) & 0.110 & 0.031 \\
IQM Garnet         & 4 & 3 ($k$)       & 0.080 & 0.027 \\
IQM Garnet         & 4 & 4 ($k\!+\!1$) & 0.090 & 0.029 \\
IQM Garnet         & 6 & 5 ($k\!-\!1$) & 0.000 & 0.000 \\
IQM Garnet         & 6 & 6 ($k$)       & 0.000 & 0.000 \\
IQM Garnet         & 6 & 7 ($k\!+\!1$) & 0.010 & 0.010 \\
\addlinespace
IonQ Forte-1       & 4 & 2 ($k\!-\!1$) & 0.180 & 0.038 \\
IonQ Forte-1       & 4 & 3 ($k$)       & 0.090 & 0.028 \\
IonQ Forte-1       & 4 & 4 ($k\!+\!1$) & 0.060 & 0.024 \\
IonQ Forte-1       & 6 & 5 ($k\!-\!1$) & 0.060 & 0.024 \\
IonQ Forte-1       & 6 & 6 ($k$)       & 0.040 & 0.019 \\
IonQ Forte-1       & 6 & 7 ($k\!+\!1$) & 0.060 & 0.024 \\
\addlinespace
Rigetti Ankaa-3    & 4 & 2 ($k\!-\!1$) & 0.030 & 0.017 \\
Rigetti Ankaa-3    & 4 & 3 ($k$)       & 0.090 & 0.028 \\
Rigetti Ankaa-3    & 4 & 4 ($k\!+\!1$) & 0.040 & 0.020 \\
Rigetti Ankaa-3    & 6 & 5 ($k\!-\!1$) & 0.030 & 0.017 \\
Rigetti Ankaa-3    & 6 & 6 ($k$)       & 0.020 & 0.014 \\
Rigetti Ankaa-3    & 6 & 7 ($k\!+\!1$) & 0.020 & 0.014 \\
\addlinespace
SV1                & 4 & 2 ($k\!-\!1$) & 0.580 & 0.049 \\
SV1                & 4 & 3 ($k$)       & 0.150 & 0.036 \\
SV1                & 4 & 4 ($k\!+\!1$) & 0.000 & 0.000 \\
SV1                & 6 & 5 ($k\!-\!1$) & 0.400 & 0.049 \\
SV1                & 6 & 6 ($k$)       & 0.270 & 0.044 \\
SV1                & 6 & 7 ($k\!+\!1$) & 0.090 & 0.029 \\
\bottomrule
\end{tabular*}
\end{table}

\subsection{Quantum Approximate Optimization Algorithm (QAOA) Results}

\begin{figure}[htbp]
\centering
\includegraphics[width=\columnwidth]{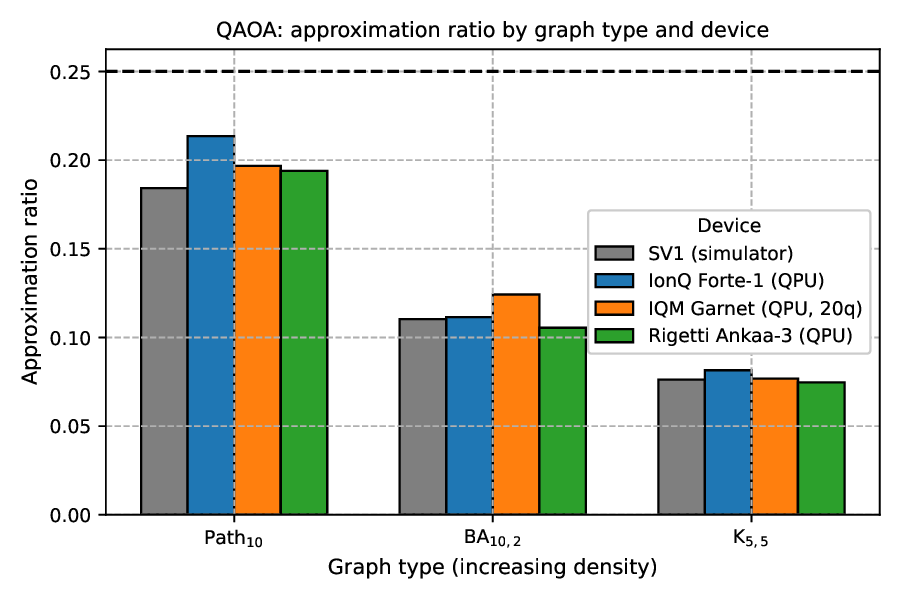}
\caption{QAOA approximation ratio by graph type (Path$_{10}$, BA$_{10,2}$, K$_{5,5}$) and device. Bars use consistent device colors, and error bars indicate uncertainty.}
\label{fig:qaoa_approx_ratio_by_graph_device}
\end{figure}

\begin{figure}[htbp]
\centering
\includegraphics[width=\columnwidth]{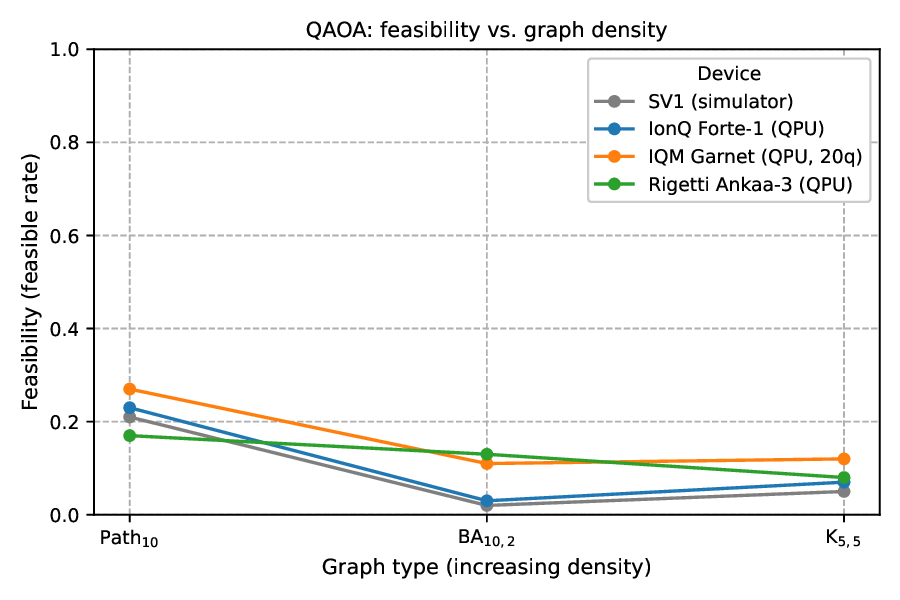}
\caption{Feasibility (valid-bitstring rate) as a function of graph density for all devices.}
\label{fig:qaoa_feasibility_vs_density}
\end{figure}

\begin{figure}[htbp]
\centering
\includegraphics[width=\columnwidth]{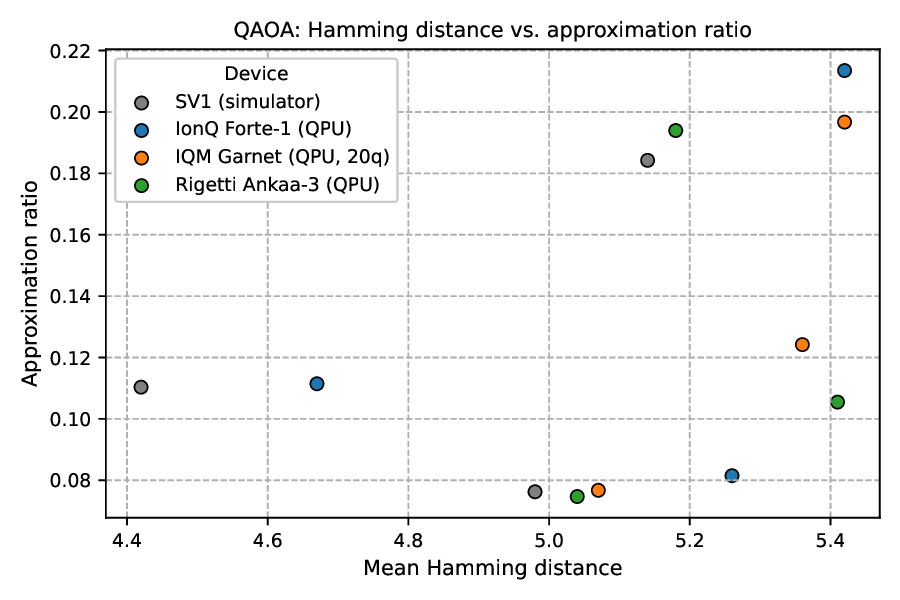}
\caption{Relationship between mean Hamming distance and approximation ratio for QAOA instances across devices.}
\label{fig:qaoa_hamming_vs_approx_ratio}
\end{figure}

The QAOA experiments targeted the Minimum Vertex Cover problem on three 10-node (see Table~\ref{tab:quantum_algorithms}) graph instances with increasing edge density: the sparse Path$_{10}$, the moderately connected BA$_{10,2}$, and the dense complete bipartite graph $K_{5,5}$. All runs used depth $p = 1$ and the measurement and optimization procedure described in the methodology. Table~\ref{tab:qaoa_metrics} provides a quantitative overview of approximation ratios, feasibility rates, success probabilities, and Hamming distances across all devices and graph types, while Figs.~\ref{fig:qaoa_approx_ratio_by_graph_device}, \ref{fig:qaoa_feasibility_vs_density}, and~\ref{fig:qaoa_hamming_vs_approx_ratio} visualize the dependence on graph density and hardware platform.

Across all backends, the achieved approximation ratios remain modest ($r \approx 0.07$–$0.21$), which is consistent with the expressivity limits of depth-$p=1$ QAOA on 10-node vertex-cover instances. SV1 generally attains the highest or near-highest ratios for each graph, while the trapped-ion and superconducting devices track similar trends with small quantitative differences. The impact of graph density is visible across all platforms: approximation ratios are slightly higher on the sparse Path$_{10}$ instance and decrease for BA$_{10,2}$ and $K_{5,5}$.

Feasibility rates and Hamming distances provide additional insight into the noise sensitivity of these shallow circuits. Denser graphs exhibit lower feasibility (a smaller fraction of valid vertex covers) and larger average Hamming distances, indicating that hardware noise pushes samples further away from the optimal bitstrings. Success probabilities for sampling an optimal solution remain low (typically a few percent) across all devices, underscoring the challenge of solving constrained combinatorial problems with very shallow QAOA in the NISQ regime. Overall, the results confirm that QAOA can capture problem structure but that its practical accuracy and robustness depend strongly on hardware fidelity, connectivity, and achievable circuit depth~\cite{b1}.

\begin{table}[h]
\caption{QAOA benchmarking metrics across devices and graph instances. Reported are the approximation ratio $r=C_{\mathrm{quantum}}/C_{\mathrm{opt}}$ with uncertainty, feasibility rate (percentage of valid bitstrings), success probability (optimal sample fraction), and average Hamming distance between measured and optimal bitstrings}
\label{tab:qaoa_metrics}
\centering
\scriptsize
\begin{tabular*}{\textwidth}{@{\extracolsep\fill}l c c c c c l}
\toprule
Device & Graph type & Approx.\ ratio & $\pm$Err. & Feas.\ (\%) & Succ. & Avg.\ Hamming dist. \\
\midrule
SV1 (Sim.)         & Path$_{10}$     & 0.184 & 16.93 & 21.0 & 0.02 & 5.14 \\
SV1 (Sim.)         & BA$_{10,2}$     & 0.110 & 20.88 & 2.0  & 0.03 & 4.42 \\
SV1 (Sim.)         & K$_{5,5}$       & 0.076 & 34.03 & 5.0  & 0.03 & 4.98 \\
\addlinespace
IonQ Forte-1       & Path$_{10}$     & 0.213 & 14.25 & 23.0 & 0.02 & 5.42 \\
IonQ Forte-1       & BA$_{10,2}$     & 0.111 & 23.21 & 3.0  & 0.02 & 4.67 \\
IonQ Forte-1       & K$_{5,5}$       & 0.081 & 39.93 & 7.0  & 0.03 & 5.26 \\
\addlinespace
IQM Garnet         & Path$_{10}$     & 0.197 & 15.58 & 27.0 & 0.02 & 5.42 \\
IQM Garnet         & BA$_{10,2}$     & 0.124 & 26.05 & 11.0 & 0.02 & 5.36 \\
IQM Garnet         & K$_{5,5}$       & 0.077 & 43.44 & 12.0 & 0.03 & 5.07 \\
\addlinespace
Rigetti Ankaa-3    & Path$_{10}$     & 0.194 & 15.41 & 17.0 & 0.02 & 5.18 \\
Rigetti Ankaa-3    & BA$_{10,2}$     & 0.105 & 31.18 & 13.0 & 0.03 & 5.41 \\
Rigetti Ankaa-3    & K$_{5,5}$       & 0.075 & 41.43 & 8.0  & 0.03 & 5.04 \\
\bottomrule
\end{tabular*}
\end{table}

\subsection{Cross-Algorithm Comparison}\label{subsec8}

When comparing all five benchmark algorithms, a consistent architectural pattern emerges across quantum hardware platforms. The trapped-ion systems (IonQ Aria-1 and Forte-1) achieve the highest fidelities, success probabilities, and approximation ratios owing to their native all-to-all connectivity, low two-qubit gate error rates, and long qubit coherence times. In contrast, superconducting architectures (IQM Garnet and Rigetti Ankaa-3) exhibit significantly faster gate execution but suffer from higher cumulative error in deeper circuits due to limited nearest-neighbor coupling and greater cross-talk sensitivity. The simulator results establish an ideal upper bound, isolating the effects of physical noise, calibration drift, and restricted connectivity.

Across all devices, algorithmic performance correlates strongly with circuit depth and entanglement complexity. Shallow-depth protocols such as the Bell and GHZ experiments consistently yield high fidelities on every device, indicating reliable few-qubit control. However, deeper and more gate-intensive algorithms—Grover’s search, the QFT round-trip, and QAOA—demonstrate pronounced degradation on physical hardware. These trends underscore the trade-off between algorithmic expressivity and noise resilience in the current NISQ regime.

\section{Discussion}\label{sec5}
The results obtained across all algorithms highlight several important characteristics of current Noisy Intermediate-Scale Quantum (NISQ) hardware.  
First, algorithmic performance is highly dependent on device architecture and connectivity.  
The trapped-ion systems demonstrated superior stability and entanglement quality due to their all-to-all qubit coupling and low gate noise, enabling accurate realization of algorithms requiring global interactions, such as GHZ and QAOA.  
In contrast, superconducting devices were limited by nearest-neighbor topologies, leading to deeper circuits with higher accumulated error, particularly in algorithms demanding long coherence and multiple entangling operations.

A clear trend emerges between circuit depth and performance degradation.  
Algorithms with shallow depth and fewer entangling gates (Bell and GHZ) showed strong agreement with theoretical expectations across all platforms, confirming that small-scale entanglement is now routine on NISQ hardware.  
However, deeper circuits such as the QFT and Grover’s search accumulated significant two-qubit gate errors, which compounded over multiple layers.  
QAOA further exposed these limitations, as its hybrid optimization loop magnifies imperfections in state preparation and measurement.  
This scaling behavior reinforces prior findings that circuit depth—not only qubit count—remains the dominant constraint for near-term quantum computation.

The comparative analysis across devices also provides insight into architectural trade-offs.  
While trapped-ion processors offer long coherence times and high gate fidelities, their slower gate speeds result in longer execution times and susceptibility to drift over extended calibration cycles.  
Superconducting devices, conversely, achieve faster gate operation but are more affected by crosstalk, limited connectivity, and calibration sensitivity.  
These contrasting characteristics suggest that future quantum platforms may benefit from heterogeneous or hybrid approaches, combining the connectivity advantages of ion traps with the scalability of superconducting systems.

From an algorithmic perspective, the experiments demonstrate that noise does not uniformly affect all computational layers.  
For instance, Grover’s amplitude amplification is especially sensitive to phase errors, while QAOA’s cost Hamiltonian suffers from stochastic parameter fluctuations that reduce convergence precision.  
Understanding these algorithm-specific sensitivities is essential for designing targeted error mitigation and optimal compilation strategies.  
Furthermore, device-aware transpilation and connectivity mapping—already crucial in QFT circuits—will become indispensable as algorithmic depth increases.

Overall, these results illustrate the fine balance between theoretical algorithm design and practical hardware constraints in the NISQ regime.  
They confirm that while idealized quantum behavior is reproducible on small systems, extending these methods to larger scales requires not only better qubits but also co-designed compilation, calibration, and noise-aware optimization techniques.  
This interplay between hardware architecture and algorithmic expressivity defines the current frontier of quantum computing research.

\section{Conclusion}\label{sec6}
This study presented a comprehensive cross-platform evaluation of five foundational quantum algorithms—Bell state preparation, GHZ state generation, Quantum Fourier Transform, Grover’s Search, and the Quantum Approximate Optimization Algorithm (QAOA)—executed on both simulated and physical quantum devices. The experiments collectively characterize the current capabilities and limitations of Noisy Intermediate-Scale Quantum (NISQ) hardware in performing representative classes of quantum tasks, including state preparation, phase estimation, amplitude amplification, and hybrid optimization.

Across all experiments, the trapped-ion architectures (IonQ Aria-1 and Forte-1) consistently achieved the highest fidelities, success probabilities, and approximation ratios. Their all-to-all connectivity and long coherence times enabled the most faithful realization of multi-qubit entanglement and variational circuits. In contrast, superconducting devices (IQM Garnet and Rigetti Ankaa-3) demonstrated competitive performance for shallow circuits but exhibited rapid fidelity loss with increasing circuit depth, reflecting higher two-qubit error rates and connectivity constraints. The simulator results served as a noise-free reference, providing ideal benchmarks for each algorithm’s theoretical limits.

Algorithmic depth and entanglement complexity emerged as key performance determinants. Shallow algorithms such as Bell and GHZ states showed robust and reproducible behavior across all devices, while deeper protocols—QFT, Grover, and QAOA—exhibited pronounced degradation due to cumulative gate and readout errors. These findings emphasize the inherent trade-off between algorithmic expressivity and hardware noise resilience, defining the practical boundaries of current NISQ systems.

Overall, the results confirm that today’s quantum processors can reliably generate and manipulate entangled states, perform small-scale Fourier and search operations, and execute hybrid optimization routines. However, their accuracy and scalability remain limited by coherence times, gate fidelity, and connectivity topology. Continued improvements in qubit quality, error correction, and architectural design will be essential to extend these results to larger problem instances and more expressive algorithmic depths. Future work will focus on increasing circuit depth \(p\) in QAOA, benchmarking additional optimization problems, and integrating error mitigation strategies to approach fault-tolerant performance.


\section*{Declarations}

\begin{itemize}
    \item \textbf{Funding:} The authors did not receive support from any organization for the submitted work.
    \item \textbf{Conflict of Interest:} The authors have no competing interests to declare that are relevant to the content of this article.
    \item \textbf{Ethics Approval:} Not applicable.
    \item \textbf{Consent to Participate:} Not applicable.
    \item \textbf{Consent for Publication:} All authors have read and approved the final manuscript for publication.
    \item \textbf{Data Availability:} The datasets generated and analyzed during the current study are available from the corresponding author on reasonable request.
    \item \textbf{Author Contributions:} \textbf{Askar Oralkhan} performed the algorithm implementation, data collection, and analysis, and wrote the first draft. \textbf{Temirlan Zhaxalykov} provided methodology refinement, supervision, and critical revision of the manuscript.
    \item \textbf{AI Disclosure:} The authors acknowledge the use of ChatGPT for language editing/polishing. The authors are fully responsible for the content and accuracy of the final manuscript.
\end{itemize}


\begin{thebibliography}{00}

\bibitem{b1} Nielsen, M.A., Chuang, I.L.: Quantum Computation and Quantum Information, 10th Anniversary ed. Cambridge University Press, Cambridge (2010)

\bibitem{b2} Preskill, J.: Quantum computing in the NISQ era and beyond. Quantum \textbf{2}, 79 (2018). \url{https://doi.org/10.22331/q-2018-08-06-79}

\bibitem{b3} Montanaro, A.: Quantum algorithms: An overview. npj Quantum Inf. \textbf{2}, 15023 (2016). \url{https://doi.org/10.1038/npjqi.2015.23}

\bibitem{b4} DiVincenzo, D.P.: The physical implementation of quantum computation. Fortschr. Phys. \textbf{48}, 771--783 (2000). \url{https://doi.org/10.1002/1521-3978(200009)48:9/11<771::AID-PROP771>3.0.CO;2-E}

\bibitem{b5} von Neumann, J.: First draft of a report on the EDVAC. Univ. Pennsylvania, Moore School of Electrical Engineering, Tech. Rep. (1945)

\bibitem{b6} Blatt, R., Wineland, D.: Entangled states of trapped atomic ions. Nature \textbf{453}, 1008--1015 (2008). \url{https://doi.org/10.1038/nature07125}

\bibitem{b7} Krantz, P., et al.: A quantum engineer's guide to superconducting qubits. Appl. Phys. Rev. \textbf{6}, 021318 (2019). \url{https://doi.org/10.1063/1.5089550}

\bibitem{b8} Cross, A.W., Bishop, L.S., Sheldon, S., Nation, P.D., Gambetta, J.M.: Validating quantum computers using randomized model circuits. Phys. Rev. A \textbf{100}, 032328 (2019). \url{https://doi.org/10.1103/PhysRevA.100.032328}

\bibitem{b9} Wright, K., et al.: Benchmarking an 11-qubit quantum computer. Nat. Commun. \textbf{10}, 5464 (2019). \url{https://doi.org/10.1038/s41467-019-13534-2}

\bibitem{b10} Arute, F., et al.: Quantum supremacy using a programmable superconducting processor. Nature \textbf{574}, 505--510 (2019). \url{https://doi.org/10.1038/s41586-019-1666-5}

\bibitem{b11} McArdle, S., Endo, S., Aspuru-Guzik, A., Benjamin, S.C., Yuan, X.: Quantum computational chemistry. Rev. Mod. Phys. \textbf{92}, 015003 (2020). \url{https://doi.org/10.1103/RevModPhys.92.015003}

\bibitem{b12} Campbell, E.T., Terhal, B.M., Vuillot, C.: Roads towards fault-tolerant universal quantum computation. Nature \textbf{549}, 172--179 (2017). \url{https://doi.org/10.1038/nature23460}

\bibitem{b13} Blinov, S., Wu, B., Monroe, C.: Comparison of cloud-based ion trap and superconducting quantum computer architectures. AVS Quantum Sci. \textbf{3}, 034101 (2021). \url{https://doi.org/10.1116/5.0058187}

\bibitem{b14} Linke, N.M., et al.: Experimental comparison of two quantum computing architectures. Proc. Natl. Acad. Sci. USA \textbf{114}, 3305--3310 (2017). \url{https://doi.org/10.1073/pnas.1618020114}

\bibitem{b15} Zhu, D., et al.: Cross-platform comparison of arbitrary quantum states. Nat. Commun. \textbf{13}, 6675 (2022). \url{https://doi.org/10.1038/s41467-022-34279-5}

\bibitem{b16} Murali, P., et al.: Full-stack, real-system quantum computer studies: Architectural comparisons and design insights. arXiv preprint arXiv:1905.11349 (2019)

\bibitem{b17} Montanez-Barrera, J.A., Michielsen, K., Bernal Neira, D.E.: Evaluating the performance of quantum processing units at large width and depth. arXiv preprint arXiv:2502.06471 (2025)

\bibitem{b18} Robertson, R., Doucet, E., Spicer, E., Deffner, S.: Simon’s algorithm in the NISQ cloud. Entropy \textbf{27}, 658 (2025). \url{https://doi.org/10.3390/e27070658}

\bibitem{b19} Schwaller, N., Vento, V., Galland, C.: Experimental QND measurements of complementarity on two-qubit states with IonQ and IBM Q quantum computers. Quantum Inf. Process. \textbf{21}, 43 (2022). \url{https://doi.org/10.1007/s11128-021-03354-z}

\bibitem{b20} IonQ, Inc.: IonQ Aria and Forte quantum systems (2025). \url{https://www.ionq.com/quantum-systems/aria}. Accessed 3 Nov 2025

\bibitem{b21} IQM Quantum Computers: IQM Garnet superconducting quantum processor (2025). \url{https://www.iqmacademy.com/qpu/garnet/}. Accessed 3 Nov 2025

\bibitem{b22} Rigetti Computing: Rigetti Ankaa-3 quantum processing unit (2025). \url{https://qcs.rigetti.com/qpus}. Accessed 3 Nov 2025

\bibitem{b23} Amazon Web Services: Amazon Braket State Vector Simulator (SV1) (2025). \url{https://docs.aws.amazon.com/braket/latest/developerguide/braket-submit-tasks-simulators.html}. Accessed 3 Nov 2025

\bibitem{b24} Bell, J.S.: On the Einstein Podolsky Rosen paradox. Physics \textbf{1}, 195--200 (1964)

\bibitem{b25} Clauser, J.F., Horne, M.A., Shimony, A., Holt, R.A.: Proposed experiment to test local hidden-variable theories. Phys. Rev. Lett. \textbf{23}, 880--884 (1969)

\bibitem{b26} Aspect, A., Grangier, P., Roger, G.: Experimental tests of realistic local theories via Bell’s theorem. Phys. Rev. Lett. \textbf{47}, 460--463 (1981)

\bibitem{b27} Friis, N., et al.: Observation of entangled states of a fully controlled 20-qubit system. Phys. Rev. X \textbf{8}, 021012 (2018). \url{https://doi.org/10.1103/PhysRevX.8.021012}

\bibitem{b28} Blatt, R., Wineland, D.: Entangled states of trapped atomic ions. Nature \textbf{453}, 1008--1015 (2008). \url{https://doi.org/10.1038/nature07125}

\bibitem{b29} Shor, P.W.: Algorithms for quantum computation: discrete logarithms and factoring. Proc. 35th Annu. Symp. Found. Comput. Sci. (FOCS), 124--134 (1994). \url{https://doi.org/10.1109/SFCS.1994.365700}

\bibitem{b30} Coppersmith, D.: An approximate Fourier transform useful in quantum factoring. IBM Research Report, RC19642 (1994)

\bibitem{b31} Nam, Y., et al.: Ground-state energy estimation of the water molecule on a trapped-ion quantum computer. npj Quantum Inf. \textbf{6}, 33 (2020). \url{https://doi.org/10.1038/s41534-020-0259-3}

\bibitem{b32} Grover, L.K.: A fast quantum mechanical algorithm for database search. Proc. 28th Annu. ACM Symp. Theory Comput. (STOC), 212--219 (1996). \url{https://doi.org/10.1145/237814.237866}

\bibitem{b33} Figgatt, C., et al.: Complete 3-qubit Grover search on a programmable quantum computer. Nat. Commun. \textbf{8}, 1918 (2017). \url{https://doi.org/10.1038/s41467-017-01904-7}

\bibitem{b34} Farhi, E., Goldstone, J., Gutmann, S.: A quantum approximate optimization algorithm. arXiv preprint arXiv:1411.4028 (2014)

\bibitem{b35} Cerezo, M., et al.: Variational quantum algorithms. Nat. Rev. Phys. \textbf{3}, 625--644 (2021). \url{https://doi.org/10.1038/s42254-021-00348-9}

\bibitem{b36} Hadfield, S., et al.: From the quantum approximate optimization algorithm to a quantum alternating operator ansatz. Algorithms \textbf{12}, 34 (2019). \url{https://doi.org/10.3390/a12020034}

\bibitem{b37} Egger, T., Mareček, J., Woerner, S.: Warm-starting quantum optimization. Quantum \textbf{5}, 479 (2021). \url{https://doi.org/10.22331/q-2021-06-17-479}

\bibitem{b38} Kandala, A., et al.: Hardware-efficient variational quantum eigensolver for small molecules and quantum magnets. Nature \textbf{549}, 242--246 (2017). \url{https://doi.org/10.1038/nature23879}

\bibitem{b39} Xanadu Quantum Technologies Inc.: PennyLane: A cross-platform Python library for differentiable quantum programming (2025). \url{https://pennylane.ai/}. Accessed 30 Nov 2025

\bibitem{b40} Amazon Web Services: Amazon Braket: Quantum computing service (2025). \url{https://aws.amazon.com/braket/}. Accessed 30 Nov 2025
\end{thebibliography}
\end{document}